\documentclass[
 aip,amsmath,amssymb,
 reprint,
]{revtex4-2}

\usepackage{comment}
\usepackage{graphicx}
\usepackage{dcolumn}
\usepackage{bm}

\usepackage[utf8]{inputenc}
\usepackage[T1]{fontenc}
\usepackage{mathptmx}
\usepackage{color} 
\usepackage{svg}
\usepackage[normalem]{ulem}
\definecolor{mcolor}{rgb}{0.0,0.5,0.0}

\begin{document}

\preprint{AIP/123-QED}

\title{Self-interaction corrected Kohn-Sham effective potentials using the density-consistent effective potential method}

\author{Carlos M. Diaz}
 \email{cmdiaz@protonmail.com}
 \affiliation{Department of Physics, University of Texas at El Paso, El Paso, TX 79968, USA}
 \affiliation{Computational Science Program, University of Texas at El Paso, El Paso, Texas 79968, USA}

\author{Luis Basurto}
\affiliation{Department of Physics, University of Texas at El Paso, El Paso, TX 79968, USA}

\author{Santosh Adhikari}
\affiliation{Department of Physics, Temple University, Philadelphia, Pennsylvania 19122, USA}

\author{Yoh Yamamoto}
\affiliation{Department of Physics, University of Texas at El Paso, El Paso, TX 79968, USA}

\author{Adrienn Ruzsinszky}
\affiliation{Department of Physics, Temple University, Philadelphia, Pennsylvania 19122, USA}

\author{Tunna Baruah}
\affiliation{Department of Physics, University of Texas at El Paso, El Paso, TX 79968, USA}
\affiliation{Computational Science Program, University of Texas at El Paso, El Paso, Texas 79968, USA}

\author{Rajendra R. Zope}
 \email{rzope@utep.edu}
\affiliation{Department of Physics, University of Texas at El Paso, El Paso, TX 79968, USA}
\affiliation{Computational Science Program, University of Texas at El Paso, El Paso, Texas 79968, USA}

\date{\today}

\begin{abstract}
 Density functional theory (DFT) and beyond-DFT methods are often used in combination with photoelectron spectroscopy to obtain physical insights into the electronic structure of molecules and solids. The Kohn-Sham eigenvalues are not electron removal energies except for the highest occupied orbital. The eigenvalues of the highest occupied molecular orbitals often underestimate the electron removal or ionization energies due to the self-interaction (SI) errors in approximate density functionals. 
  In this work, we adapt and implement the density-consistent effective potential (DCEP) method of Kohut, Ryabinkin, and Staroverov to obtain SI corrected local effective potentials from the SI corrected Fermi-L\"owdin  orbitals and density in the FLOSIC scheme. The implementation is used to obtain the density of states (photoelectron spectra) and HOMO-LUMO gaps for a set of molecules and polyacenes.  Good agreement with experimental values is obtained compared to  a range of SI uncorrected density functional approximations. 
\end{abstract}

\maketitle

\section{Introduction} 
Density functional theory\cite{PhysRev.140.A1133,RevModPhys.61.689,jones2015density} (DFT) and beyond-DFT methods are often used in combination with photoelectron spectroscopy to obtain physical insights into the electronic structure of molecules and solids.
The Kohn-Sham (KS) eigenvalues are not electron removal energies except for the highest occupied one\cite{almbladh1985,PhysRevA.30.2745,Perdew1982,Perdew1997,PhysRevB.60.4545}, but they often, though not  always, provide good approximations to electron binding energies (EBEs)\cite{PhysRevB.70.134422,C8SC03862G,PhysRevB.79.201205,PhysRevB.73.205407}. Eigenvalues of the range separated hybrid functionals\cite{doi:10.1063/1.1383587} using tuned separation parameter generally provide good approximations to EBEs due to mitigation of self-interaction (SI) errors\cite{doi:10.1063/1.1383587,baer2010tuned,kronik2012excitation}.
SI error has a large role in the underestimation of the magnitude of the KS eigenvalues. For excitation energies, the HOMO-LUMO orbital difference in time-dependent KS theory plays a significant role.  
The first ionization energy and electron affinity in the exact KS theory are determined by the HOMO (highest partly-occupied molecular orbital) energies of the system before and after the addition of a fraction of an electron. The HOMO-LUMO gap at fixed electron number is not equal to the first-excitation energy, but is close enough to it in a molecule to enable an accurate calculation in TDDFT (time-dependent density functional theory) of the charge transfer excitation of donor and acceptor molecules. 
The standard implementations of self-interaction correction methods in the generalized Kohn-Sham scheme\cite{PhysRevB.28.5992, PhysRevA.95.052505} correct only the occupied orbitals while the unoccupied orbitals see only the mean-field DFA potential. We here seek a way to make the  unoccupied orbitals (including the Rydberg states) see a self-interaction-corrected potential. For that purpose, we adapt and implement the the density-consistent effective potential (DCEP) method of Kohut, Ryabinkin, and Staroverov\cite{Ryabinkin2015,Kohut2014} to obtain effective local potentials from the Perdew-Zunger\cite{PhysRevB.23.5048} (PZ) self-interaction corrected orbitals and density.

The Perdew-Zunger self-interaction correction (PZSIC) approach is closer to the self-interaction correction schemes\cite{lindgren1971statistical,PhysRevA.15.2135,PhysRevB.23.5048} that remove self-interaction error on an orbital-by-orbital basis. The PZ energy results in an energy functional that is orbital dependent.
In the KS-DFT scheme, implementing such orbital-dependent functionals requires computing the multiplicative potential $v_{XC} (\vec{r})=\frac{\delta E_{XC}}{\delta \rho (\vec{r})}$ . Here $E_{XC}$ is the exchange-correlation functional and $\rho (\vec{r})$ is the total electron density. 
When the functional $E_{XC}$ is not an explicit functional of the density, the potential cannot be obtained by a straightforward evaluation. 
Traditionally, an effective potential has been obtained by solving the optimized effective potential (OEP) integral equation\cite{Sharp1953,Talman1976,PhysRevA.46.5453}. The OEP approach is often not well suited to routine calculations due to numerical instabilities and difficulties solving it in finite basis sets\cite{ivanov2002finite,yang2002direct,heaton2007optimized,kummel2008orbital,hesselmann2007numerically}.
The PZSIC method has also been implemented within the Krieger-Li-Iafrate (KLI) approximation\cite{doi:10.1063/1.481421, doi:10.1063/1.1370527, Tong1997, messud2008improved, PhysRevB.77.121204, PhysRevA.103.042811} and within OEP using the real space approach \cite{Korzdorfer2008}. But as mentioned earlier, OEP implementations using finite Gaussian basis sets are usually fraught with numerical instabilities.  
The recently developed method by Kohut, Ryabinkin, and Staroverov provides a practical alternative to the OEP method to obtain KS potentials in a straightforward manner\cite{Ryabinkin2015,Kohut2014,Ospadov2017,PhysRevLett.111.013001}.
Kohut, Ryabinkin, and Staroverov in Ref. [\onlinecite{Kohut2014}] have laid out a hierarchy of successively more accurate approximations, ending in DCEP. 
In this work, we adapt the DCEP method and apply it  to the PZSIC using the Fermi-L\"owdin orbital self-interaction correction scheme\cite{doi:10.1063/1.4907592} (FLOSIC) to obtain the self-interaction corrected eigenvalues and corresponding orbitals.
The density of states (photoelectron spectra) and HOMO-LUMO gaps obtained using this approach are compared with  experimental values for several molecules.

In Sec. \ref{sec:rks} we describe the DCEP method and validate our implementation in the UTEP-NRLMOL code\cite{UTEPNRLMOL}. 
In Sec. \ref{sec:rks_flosic} we present our adaptation of the DCEP to the FLOSIC method. The computational details and results are presented in Sec.~\ref{sec:results}.

\section{Ryabinkin-Kohut-Staroverov Method}\label{sec:rks}
The simplest approximation to the OEP relies on the idea developed by Slater\cite{Slater1951} to construct an orbital-averaged potential weighted by $\vert\phi_i \vert^2/\rho $ where $\vert \phi_i \vert^2$ and $\rho$ are respectively density of the $i^{th}$ orbital and total electron density. In the context of Hartree-Fock (HF) approximation, this results in the so-called Slater potential, 
\begin{equation}
    v_S(\vec{r}) 
    = \frac{1}{\rho(\vec{r})}\sum_{i=1}^N \phi_i^*\hat{K}\phi_i =-\frac{1}{2\rho(\vec{r})} \int\frac{|\gamma((\vec{r},\vec{r}\,')|^2}{\vert\vec{r}-\vec{r}\,'\vert} d\vec{r}\,'
\end{equation}
with the Fock exchange operator $\hat{K}$ and reduced density matrix $\gamma(\vec{r},\vec{r}\,')=\sum_i^N\phi_i(\vec{r})\phi_i^*(\vec{r}\,')$, where $N$ is the number of occupied orbitals.
Kohut \textit{et al.}\cite{Kohut2014} in 2014 defined a methodology to obtain higher-order approximations to the OEP for Hartree-Fock calculations.  The method begins by rearranging the Fock equations as given by
\begin{equation}
    \Bigg[ -\frac{1}{2} \nabla^{2} + v_{ext}(\vec{r}) + v_{H}^{HF}(\vec{r}) + \hat{K} \Bigg] \phi_i^{HF} = \epsilon _i^{HF} \phi_i^{HF} .
\end{equation}
Multiplying both sides in the above equation by $\phi_i^{HF}$, summing over $i$ from 1 to $N$, and then dividing both sides by $\rho^{HF}(\vec{r})$ gives
\begin{equation}\label{eq:I_HF}
    \frac{\tau_L^{HF}(\vec{r})}{\rho^{HF}(\vec{r})} + v_{ext}(\vec{r}) + v_{H}^{HF}(\vec{r}) + v_S^{HF}(\vec{r}) = -\overline{I}^{HF}(\vec{r})   
\end{equation}
where $\tau_L^{HF}(\vec{r})$ is the HF kinetic energy density in Laplacian form and $\bar{I}^{HF}(\vec{r})$ is the HF average local ionization energy.
Likewise, repeating these steps for the  KS equations gives
\begin{equation}\label{eq:I_KS}
    \frac{\tau_L^{KS}(\vec{r})}{\rho^{KS}(\vec{r})} + v_{ext}(\vec{r}) + v_{H}^{KS}(\vec{r}) + v_X(\vec{r}) = -\overline{I}^{KS}(\vec{r}) .  
\end{equation}
Here, the Laplacian kinetic energy density for a given wavefunction (WF) (e.g. HF or KS) is
\begin{equation}
    \tau_L^{WF}(\vec{r}) = -\frac{1}{2}\sum_{i=1}^N \phi_i^{WF*}\nabla^2 \phi_i^{WF}
\end{equation}
and the average local ionization energy (sometimes $\overline{\epsilon}^{WF}$) is defined by
\begin{equation}
    \overline{I}^{WF}(\vec{r}) = -\frac{1}{\rho^{WF}(\vec{r})} \sum_{i=1}^N \epsilon_i^{WF} \vert \phi_i^{WF}\vert^2 .
\end{equation}
The highest level of approximations defined by Kohut, Ryabinkin, and Staroverov is DCEP.
This approximation relies on the assumption that the ground state densities in two schemes are equal. This equates to imposing the constraint $\rho^{KS}(\vec{r}) = \rho^{HF}(\vec{r})$. Subtracting Eq. (\ref{eq:I_KS}) from Eq. (\ref{eq:I_HF}) leads to 
\begin{equation}\label{eq:dcep}
    v_{X}^{DCEP}(\vec{r}) =v_{S}^{HF}(\vec{r}) + \overline{I}^{HF} - \overline{I}^{KS} + \frac{\tau_{HF}(\vec{r})}{\rho_{HF}(\vec{r})} - \frac{\tau_{KS}(\vec{r})}{\rho_{KS} (\vec{r})}
\end{equation}
where $\tau$ is the positive-definite form of the kinetic energy density, such that $ \tau^{HF}(\vec{r}) = \frac{1}{2} \sum_{i=1}^N \vert\nabla^2\phi_i^{HF}\vert^2 $.

In the DCEP method, the eigenvalues are then shifted so that $\epsilon_{HOMO}^{KS} = \epsilon_{HOMO}^{HF}$ to ensure the correct behavior
of the asymptotic potential in the limit of $r\rightarrow \infty$.
A full self-consistent calculation is performed subsequently.  In this work we define the tolerance for self-consistent calculations such that self-consistency is reached when relative difference between the potentials in the two successive iterations is less than  $10^{-8}$, that is, $\frac{||V_{n}-V_{n-1}||}{||V_n||} <10^{-8}.$

\subsection{DCEP-Hartree-Fock results}
\begin{figure}[h]
    \includegraphics[width=\columnwidth]{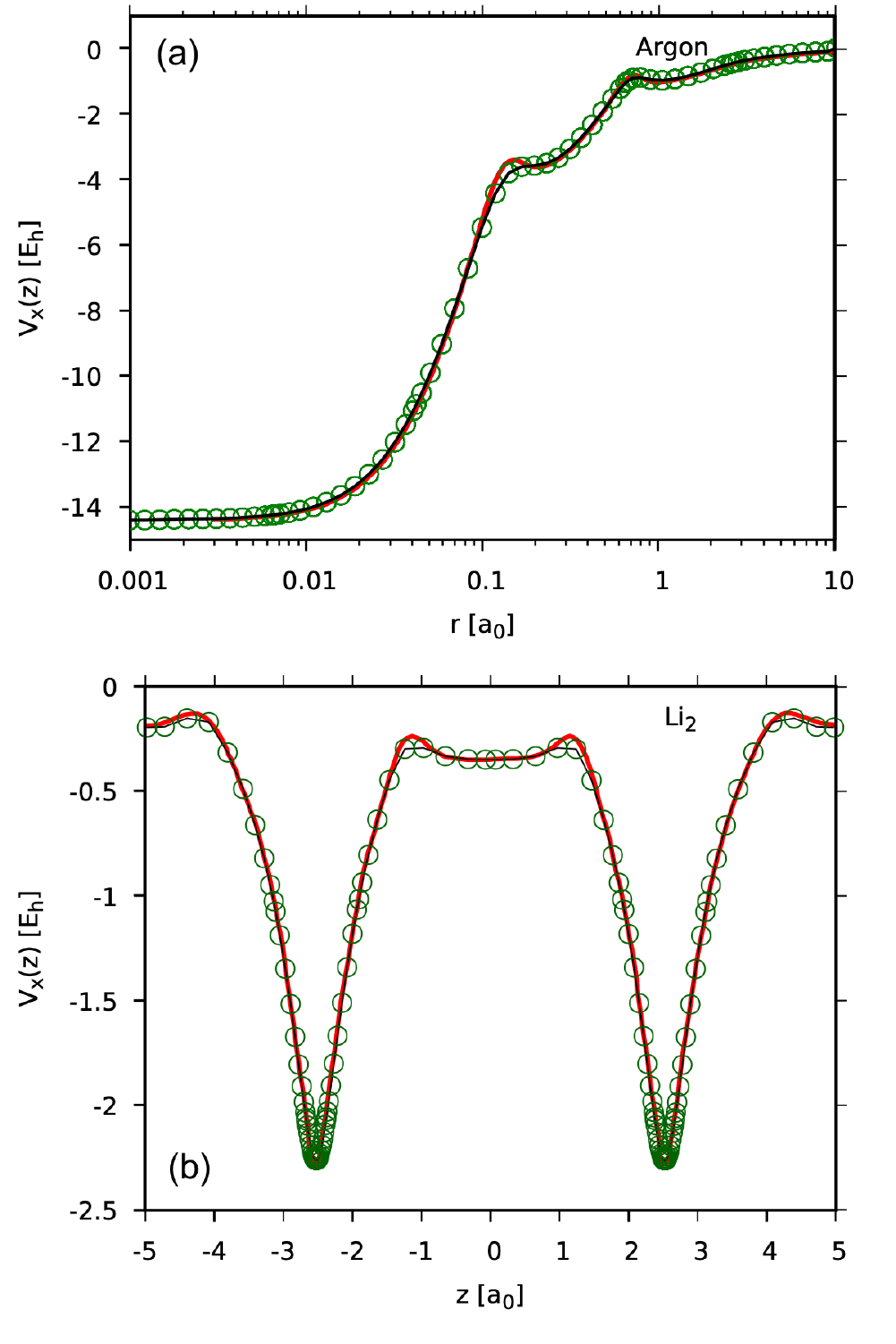}
    \caption{\label{fig:HF-Vxc} 
    The exchange potential curves obtained from the Hartree-Fock densities for (a) the argon atom plotted radially and (b) Li$_2$ plotted along the molecular axis. 
    The red curves are OEP from Ref.~\onlinecite{Kohut2014}, and the black curves are obtained from the implementation in this work. The green circles represent grid points where the data points are calculated.}
\end{figure}

We validate our DCEP implementation in the UTEP-NRLMOL code\cite{UTEPNRLMOL} by comparing our results against the 
HF results from the original DCEP work.\cite{Kohut2014} 
The NRLMOL code\cite{PhysRevB.41.7453,PhysRevB.42.3276,Pederson2000}, on which UTEP-NRLMOL is based, is a pure density functional code. The Hartree-Fock exchange was introduced in UTEP-NRLMOL code using a semi-analytic scheme similar to what is used for calculating the Coulomb energy.\cite{UTEPHF}
Fig.~\ref{fig:HF-Vxc}  shows the DCEP potentials for the Ar atom and Li$_2$ molecule obtained from the HF densities using the  NRLMOL basis set \cite{PhysRevA.60.2840}.
These potentials compare very well with the the OEP exchange potentials reported by Kohut and coworkers\cite{Kohut2014} thus validating the present implementation of the DCEP method. There is a slight difference around 0.1 a.u. where the first bump in the exchange potential is somewhat less conspicuous in the present result. We find that this is a basis set artifact and uncontracting the basis reproduces the bump in the exchange potential closely.

\section{DCEP in FLOSIC}\label{sec:rks_flosic}
We propose to extend the DCEP method to obtain a multiplicative potential from PZSIC orbitals and density.  To do so, we start with the PZSIC equations,
\begin{equation}\label{eq:pzsic}
    \Bigg[ -\frac{1}{2} \nabla^{2} + v_{ext}(\vec{r}) + v_{H}(\vec{r}) + v_{XC}(\vec{r}) + V^{i}_{SIC}(\vec{r}) \Bigg] \phi_i= \epsilon _i \phi_i.
\end{equation}
Here $i$ is the orbital index, $v_{ext}$ is the external potential, $v_H$ is the Coulomb potential and $V^{i}_{SIC}$ is the self-interaction-correction term for the $i$th orbital. $V^{i}_{SIC}$ consists of the  self-Coulomb potential $ U[\rho_i]$ term and the self-exchange-correlation potential $v_{xc}[\rho_i,0]$ term as follows, 
\begin{eqnarray}
  V^{i}_{SIC}& = & -  \{\, U[\rho_i] + v_{xc}[\rho_i,0]  \,\} \\
            & = & -\int d^3r\,' \frac{\rho_i(\vec{r}\,')}{\vert \vec{r} - \vec{r}\,'\vert}
   - v_{xc}[\rho_i,0].
\end{eqnarray}
In our case, we use the Fermi L\"owdin orbitals  (FLOs) implementation of PZSIC\cite{PEDERSON2015153,PhysRevA.95.052505} to determine $V_{SIC}^i$ such that $\phi_i = \phi_i^{FLO}$ and $\rho_i = \rho_i^{FLO} = |\phi_i^{FLO}|^2$. To form FLOs, first, a set of Fermi orbitals $F_i$ is constructed with the density matrix and $3N$ positions called Fermi orbital descriptor (FOD) positions given as
\begin{equation}
    F_i(\vec{r})=\frac{\sum_j^{N}\psi_j(a_i)\psi_j(\vec{r})}{\sqrt{\rho_i(a_i)}}.
\end{equation}
Then, the set of $F_i$'s  is orthogonalized using L\"owdin's scheme to obtain FLOs. 
Similarly, as with the previous case in Sec.~\ref{sec:rks}, we multiply both sides of Eq.~(\ref{eq:pzsic}) with $\phi_i$, sum over $i$ from 1 to $N$, and divide both sides with $\rho_{SIC}(\vec{r})$. This yields averaged-over quantities shown as
\begin{equation}\label{eq:I_SIC}
    \frac{\tau(\vec{r})}{\rho_{SIC}(\vec{r})} + v_{ext}(\vec{r}) + v_{H}(\vec{r}) + v_{XC}(\vec{r}) + \overline{V}_{SIC}(\vec{r}) 
    = -\overline{I}^{SIC}(\vec{r})
\end{equation}
where $\rho_{SIC}=\sum_i \rho_i$. Here $\overline{V}_{SIC}$ averages over the orbital SIC potentials using
\begin{equation}\label{eq:avg_sicv}
    \overline{V}_{SIC}(\vec{r}) = \sum_{i=1}^{N} v_{SIC}^{i}(\vec{r}) \frac{\rho_i(\vec{r})}{\rho_{SIC}(\vec{r})}.
\end{equation}
Finally, subtracting Eq. (\ref{eq:I_KS}) from Eq. (\ref{eq:I_SIC}) subject to the constraint in Eq. (\ref{eq:dcep}), we obtain the density-consistent effective potential 
\begin{equation}
    v_{XC}^{DCEP}(\vec{r}) =v_{XC}^{SIC}(\vec{r}) + \overline{V}_{SIC}(\vec{r}) + \overline{I}^{SIC} - \overline{I}^{KS} + \frac{\tau_{SIC}(\vec{r})}{\rho_{SIC}(\vec{r})} - \frac{\tau_{KS}(\vec{r})}{\rho_{KS} (\vec{r})}.
\end{equation}
These SIC terms are obtained from a self-consistent FLOSIC calculation with optimized sets of FODs. Once they are determined, a self-consistent calculation can be performed to obtain the SIC effective potential.

\section{Results}\label{sec:results}
\subsection{Computational Details}
The DCEP method was implemented in a highly-scalable version of the UTEP-NRLMOL\cite{UTEPNRLMOL} code developed at UTEP. The code uses a Gaussian orbital basis\cite{PhysRevA.60.2840} and an accurate numerical integration grid scheme\cite{PhysRevB.41.7453}. SI-corrected inputs were obtained from the FLOSIC code\cite{FLOSICcode,FLOSICcodep}, which is based on the NRLMOL code. The default grid in the FLOSIC code requires a much higher grid density than standard DFT calculations. The mesh generated with the FLOSIC code was reused for the DCEP method for each system. All calculations use the default NRLMOL\cite{PhysRevA.60.2840} basis set and the PBE exchange-correlation functional\cite{PhysRevLett.77.3865,PhysRevLett.78.1396}.

\subsection{Eigenvalues in DCEP-SIC}
\begin{figure*}[hb]
   \includegraphics[width=1.8\columnwidth]{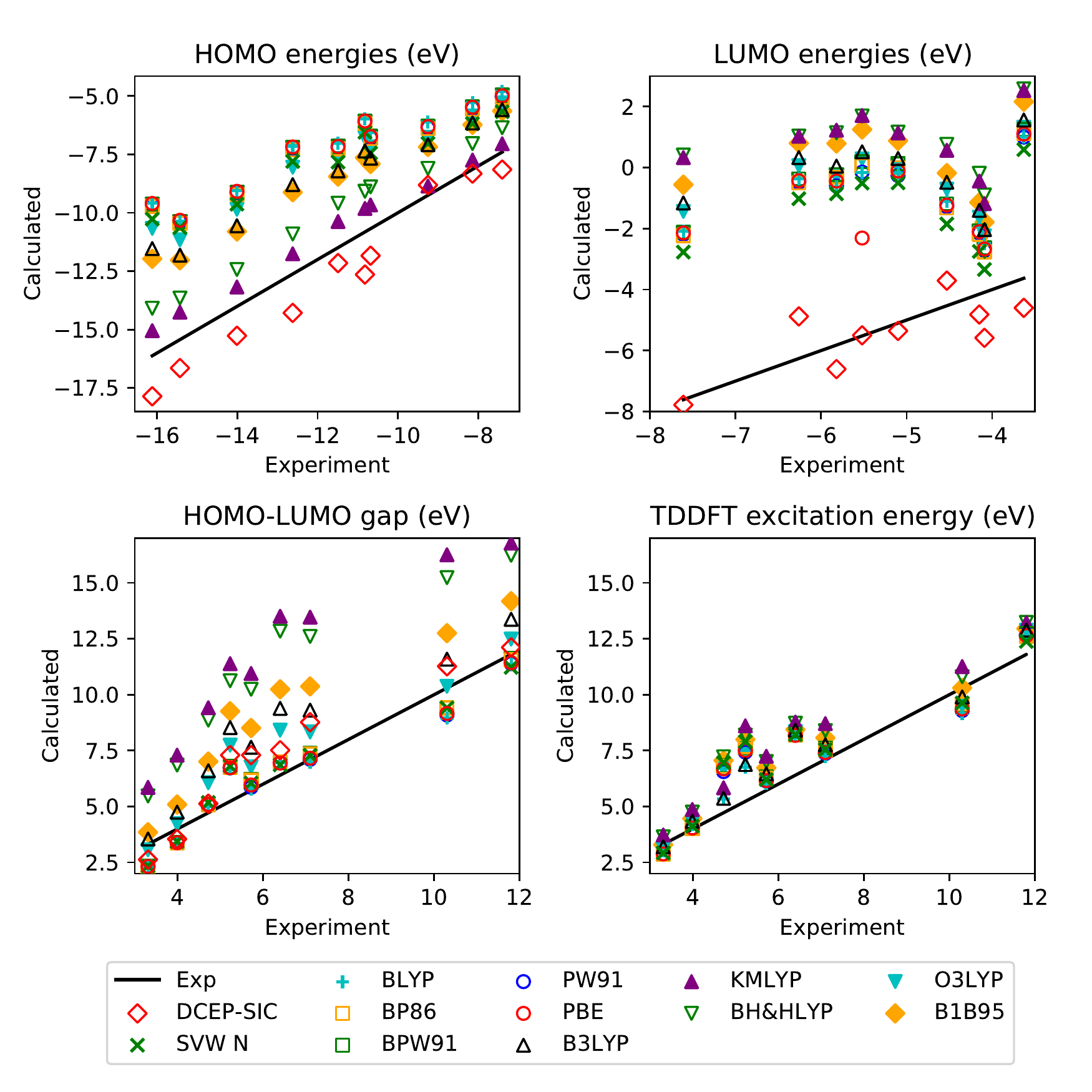} 
    \caption{\label{fig:evals} The calculated DCEP-SIC HOMO, LUMO, and HOMO-LUMO gaps (in eV) against experimental IPs and lowest excitation energies: (top left) HOMO energies, (top right) LUMO energies, (bottom left) HOMO-LUMO gaps, and (bottom right) TDDFT excitation energies.
    For comparison, values for the other functionals from Ref.~\onlinecite{Zhang2007} are shown.
    The solid line indicates the ideal agreement with experiments.}
\end{figure*}

In the FLOSIC scheme, self-consistency is obtained using Jacobi-like iterative scheme\cite{PhysRevA.95.052505} and only the occupied orbitals are affected directly by SIC. 
The DCEP method allows us to examine the effect of SI-corrected wavefunctions on unoccupied states. 
A study by Zhang and Musgrave\cite{Zhang2007} compares the HOMO, LUMO, and HOMO-LUMO gap of 11 functionals and compares those with experimental ionization potential (IP) and the lowest excitation energy for a set of molecules. 
Following Zhang and Musgrave, we define the experimental HOMO energy as the negative of the experimental ionization energy and the experimental LUMO energy as the difference between the experimental lowest excitation energy and the HOMO energy.
All the DFT and TDDFT (with adiabatic kernels) in our tables and figures other than the DCEP-SIC-PBE are from Zhang and Musgrave. 
Allen and Tozer\cite{allen2002eigenvalues} utilized the procedure of Zhao, Morrison, and Parr\cite{zhao1994electron} to determine the KS eigenvalues and HOMO-LUMO gaps from coupled-cluster BD (Brueckner Doubles) electron densities.
For the sake of clarity, we mention that our goal is not to assess various  exchange-correlation functionals, which now run into several hundreds, but to test how good the DCEP-SIC unoccupied orbitals are. For this purpose we restricted the comparison with the results reported by  Zhang and Musgrave\cite{Zhang2007} and Allen and Tozer\cite{allen2002eigenvalues}.
We have chosen C$_2$H$_2$, CO, H$_2$, HF, C$_6$H$_6$, C$_{10}$H$_8$, C$_{14}$H$_{10}$, H$_2$O, and NH$_3$ and C$_2$H$_4$ molecules. The HOMOs, LUMOs, and the HOMO-LUMO gaps calculated with DCEP-SIC are compared with the experimental values and with the results from Ref. \onlinecite{Zhang2007} in Fig. \ref{fig:evals}. The mean absolute errors (MAEs) of DCEP-SIC method for the HOMO, LUMO, and HOMO-LUMO gaps are compared with the MAEs of the 11 functionals used in Ref. \onlinecite{Zhang2007} in Table \ref{tab:homolumo}.  
We find that the DCEP-SIC eigenvalues perform well for all three categories of HOMOs, LUMOs and HOMO-LUMO gaps. The HOMO eigenvalues of the FLOSIC scheme have previously shown good agreement with vertical ionization potentials from CCSD(T) and experiments.\cite{waterpolarizability,doi:10.1063/1.5120532,C9CP06106A,doi:10.1063/5.0041265,doi:10.1063/5.0024776,PhysRevA.103.042811} 
As detailed in Sec. \ref{sec:rks}, the eigenvalues in DCEP-SIC calculations are shifted such that the DCEP HOMO matches with the FLOSIC  HOMO values.
Thus, the HOMO levels in DCEP-SIC are same as those in the FLOSIC method. 
For the chosen systems, the mean absolute error (MAE) of the FLOSIC HOMO eigenvalues is 1.09 eV with respect to the experimental ionization energies.
This MAE is relatively small when compared with the 11 functionals, with only the KMLYP functional which has about 55.7\% of Hartree-Fock exchange providing better HOMO prediction (MAE, 0.83 eV) than DCEP-SIC (FLOSIC HOMO).

Zhang and Musgrave\cite{Zhang2007} found that all 11 functionals provide rather poor estimate of the {\it experimental} LUMO eigenvalues.
We find that the DCEP-SIC provides good agreement with experiment, with a MAE of 0.73 eV. Although LDA and GGA functionals do not perform well for HOMOs and LUMOs, they all give much better estimate of HOMO-LUMO gaps when compared with the experimental lowest excitation energies. 
The  MAEs for the HOMO-LUMO gaps for these functionals range from 0.64-0.67 eV. These functionals benefit from error cancellation while taking the difference between the HOMO and LUMO energies.
In contrast the errors are much larger for the  hybrid functionals. 
The hybrid functionals are typically implemented in generalized Kohn-Sham scheme. 
The mixing of Hartree-Fock exchange with DFA in hybrid functionals improve occupied eigenvalues due to mitigation of self-interaction errors and as result HOMO eigenvalues are more accurate in hybrid functionals but the LUMO eigenvalues remain poor.
The MAE for the HOMO-LUMO gaps  for the hybrid functionals range from 1.04-5.15 eV.
 The present DCEP-SIC method performs much better with an MAE of 1.01 eV.
Zhang and Musgrave have also compared the HOMO-LUMO gaps with (the lowest) TDDFT excitation energies.
In the bottom right plot of Fig. \ref{fig:evals}, we show the comparison of the TDDFT excitation energies reported by Zhang and Musgrave with experiment to facilitate a comparison of the HOMO-LUMO gaps shown in the bottom left panel against the TDDFT excitation energies.
The plots show the DCEP-SIC HOMO-LUMO gaps compare well to the TDDFT  excitation energies (cf. Table~\ref{tab:homolumo}) for these molecules. 
Overall, the results show that the functionals that perform best for HOMO eigenvalues (KMLYP, BH, and B1B95) perform the worst for gaps since their LUMO predictions are rather poor. 
At the same time, the functionals that perform best for gaps such as PBE and BLYP are among the worst for predicting HOMO eigenvalues accurately. In contrast, the DCEP-SIC method gives reliable results for all three quantities. It, by construction, retains the accuracy of the PZSIC HOMO eigenvalues and yields accurate LUMO eigenvalues and thereby yields accurate HOMO-LUMO gaps, as borne out by the comparison of MAEs in Table~\ref{tab:homolumo}.

\begin{table}
\caption{\label{tab:homolumo}Mean absolute errors (in eV) of calculated HOMO, LUMO, HOMO-LUMO gaps, and TDDFT excitation energies with respect to experimental ionization potentials and the first excitation energies.  
}
\begin{ruledtabular}
\begin{tabular}{ccccc}
 Method &  HOMO & LUMO & HOMO-LUMO & TDDFT\\
\hline
DCEP-SIC-PBE             & 1.09 & 0.73 & 1.01 & ---\\
SVWN$^\text{*}$    & 3.69 & 3.74 & 0.67 & 1.05\\
BLYP$^\text{*}$     & 4.41 & 4.30 & 0.66 & 0.76\\
BP86$^\text{*}$     & 4.20 & 4.29 & 0.66 & 1.03\\
BPW91$^\text{*}$    & 4.30 & 4.41 & 0.66 & 1.07\\
PW91$^\text{*}$     & 4.24 & 4.21 & 0.64 & 0.99\\
PBE$^\text{*}$      & 4.30 & 4.04 & 0.65 & 1.00\\
B3LYP$^\text{*}$    & 3.13 & 4.93 & 1.80 & 0.85\\
KMLYP$^\text{*}$    & 0.83 & 5.96 & 5.15 & 1.52\\
BH\&HLYP$^\text{*}$ & 1.58 & 6.04 & 4.49 & 1.49\\
O3LYP$^\text{*}$    & 3.67 & 4.67 & 1.04 & 1.05\\
B1B95$^\text{*}$    & 2.90 & 5.44 & 2.52 & 1.20\\
\end{tabular}
\end{ruledtabular}
\begin{flushleft}
$^\text{*}$Errors obtained using the data provided in reference~\onlinecite{Zhang2007}.
\end{flushleft}
\end{table}

\subsection{Approximate photoelectron spectra of polyacenes}

\begin{figure*}[hb]
     \includegraphics[width=1.8\columnwidth]{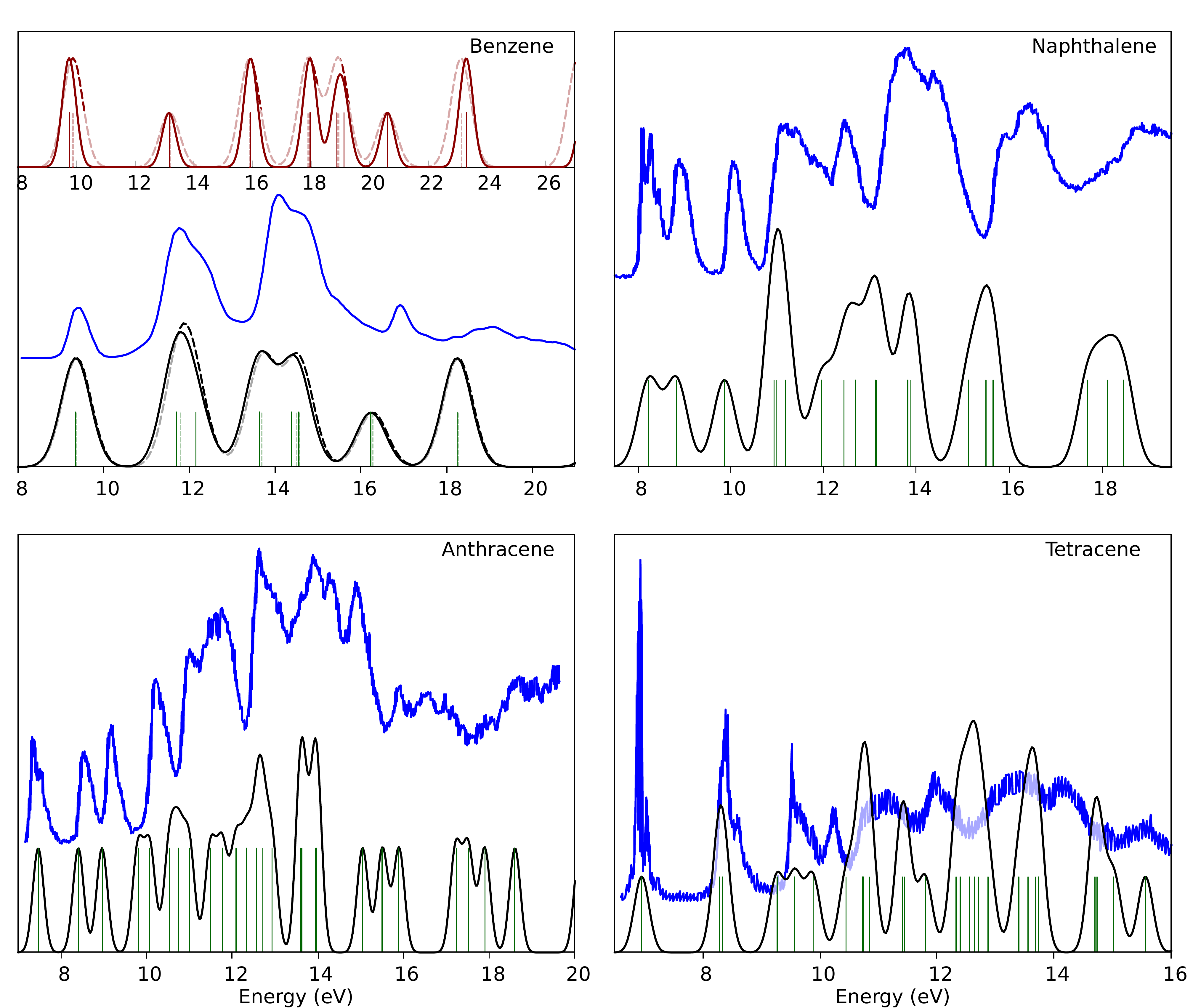}
    \caption{\label{fig:spectra} Calculated spectra of polyacenes: (top left) benzene, (top right) naphthalene, (bottom left) anthracene, and (bottom right) tetracene. The calculated spectra are shifted to match the HOMO eigenvalue with the IP of experiments. The DCEP-SIC-PBE spectra are shown in solid black and DCEP-SIC-LDA results are shown in dashed lines.
    Experimental UPS results from Refs.~\onlinecite{Yamauchi1998,Liu2011} are shown in blue. For benzene, SIC results with PBE are shown at the top of the plot in solid red and SIC-LDA results are shown using dashed red lines.}
\end{figure*}

We also examine the effect on all occupied eigenvalues by comparing the SIC and DCEP-SIC results to the photoelectron spectra of polyacenes obtained through Ultraviolet Photoelectron Spectroscopy (UPS).\cite{Yamauchi1998,Liu2011} 
The SIC eigenvalues, with LDA and PBE,  shown in the top left plot for benzene in Fig. \ref{fig:spectra} are in qualitative agreement with experimental values, but on a much broader energy scale. 
Additionally, SIC introduces an additional peak around 13.0 eV not seen in the experimental spectra.
DCEP-SIC compresses the eigenvalue spectrum and provides a much closer fit to experimental results in terms of the number of peaks and energy spacing between the eigenvalues. 
The LDA calculations result in slightly higher energies in comparison to PBE but retain the similar spacing for both SIC and DCEP-SIC. 
Similar behavior for SIC against DCEP-SIC and DCEP-SIC-LDA against DCEP-SIC-PBE eigenvalues are found also for  the other three polyacenes (naphthalene, antracene, and tetracene) and so only experimental and DCEP-SIC-PBE results  are plotted in Fig. \ref{fig:spectra}. For the larger polyacenes, DCEP-SIC shows good agreement with experimental spectra for low to mid-level energy states.  For anthracene, the spectrum displays a rigid shift higher than the experiment spectrum. In all cases, DCEP-SIC over-compresses the spectra so high-energy (core) states are underestimated.

\section{Conclusion}
We have implemented the DCEP method to generate multiplicative effective potentials from FLOSIC orbitals and densities. 
While the FLOSIC method  (within the generalized Kohn-Sham scheme) can provide accurate HOMO energies due to explicit removal of self-interaction error, the unoccupied states are essentially the same as those of the uncorrected functional. 
The same behavior can be seen for many hybrid functionals as well.
The use of a multiplicative effective potential results in much improved description of properties related to unoccupied orbitals and so gives an accurate description of the HOMO and LUMO as well as the HOMO-LUMO gap, whereas the 11 functionals tested by Zhang and Musgrave fail for one or more of these quantities.
We also find the HOMO-LUMO gap of DCEP-SIC to be comparable to TDDFT  excitation energies of GGAs and hybrid functionals.
The present results show that the DCEP-SIC  eigenvalues  provide a much better description of the experimental spectroscopic results compared to the standard PZSIC eigenvalues obtained with the FLOSIC formalism. For a set of polyacenes, we find that the DCEP-SIC eigenvalues provide a better approximation of the photoelectron spectra than the standard FLOSIC eigenvalues as 
DCEP-SIC corrects the broadened spectra of standard FLOSIC calculations.

\section*{Data Availability Statement}
The data that supports the findings of this study are available within the article.   

\begin{acknowledgments}
Authors acknowledge Dr. Po-Hao Chang for discussions. 
This work was supported by U.S. Department of Energy, Office of Science, Office of Basic Energy Sciences under Award No. DE-SC0018331.
Computational time at Texas Advanced Computing Center through NSF Grant No. TG-DMR090071 and NERSC is gratefully acknowledged.
\end{acknowledgments}

\section*{References}
\bibliography{reference}

\begin{thebibliography}{62}
\bibitem [{\citenamefont {Diaz}\ \emph {et~al.}()\citenamefont {Diaz},
  \citenamefont {Basurto}, \citenamefont {Yamamoto}, \citenamefont {Baruah},\
  and\ \citenamefont {Zope}}]{UTEPNRLMOL}%
  \BibitemOpen
  \bibfield  {author} {\bibinfo {author} {\bibfnamefont {C.~M.}\ \bibnamefont
  {Diaz}}, \bibinfo {author} {\bibfnamefont {L.}~\bibnamefont {Basurto}},
  \bibinfo {author} {\bibfnamefont {Y.}~\bibnamefont {Yamamoto}}, \bibinfo
  {author} {\bibfnamefont {T.}~\bibnamefont {Baruah}},\ and\ \bibinfo {author}
  {\bibfnamefont {R.~R.}\ \bibnamefont {Zope}},\ }\bibfield  {title} {\enquote
  {\bibinfo {title} {{UTEP-NRLMOL code: This is a Gaussian based code based on
  the old NRLMOL code that has several new features, including Hartree-Fock
  exchange used for the validation of the Ryabinkin-Kohut-Staroverov method
  presented in this work. This scalable code, developed with an emphasis on
  being optimized for memory, can simulate systems containing several thousand
  basis functions}},}\ }\href@noop {} {\ }\BibitemShut {NoStop}%
\end{thebibliography}

\end{document}